\documentclass[12pt,a4paper]{article}
\usepackage{a4wide}
\usepackage{latexsym}
\usepackage{epsf}
\usepackage{amssymb}
\begin{document}
\def\be{\begin{equation}}
\def\ee{\end{equation}}
\def\bea{\begin{eqnarray}}
\def\eea{\end{eqnarray}}

\def\pd{\partial}
\def\a{\alpha} 
\def\b{\beta}
\def\g{\gamma}
\def\d{\delta}
\def\m{\mu}
\def\n{\nu}
\def\t{\tau}
\def\l{\lambda}

\def\s{\sigma}
\def\e{\epsilon}
\def\scri{\mathcal{J}}
\def\cM{\mathcal{M}}
\def\tcM{\tilde{\mathcal{M}}}
\def\RR{\mathbb{R}}

\hyphenation{re-pa-ra-me-tri-za-tion}
\hyphenation{trans-for-ma-tions}


\begin{flushright}
IFT-UAM/CSIC-00-01\\
hep-th/0001016\\
\end{flushright}

\vspace{1cm}

\begin{center}

{\bf\Large A Comment on the Holographic Renormalization Group and
 the Soft Dilaton Theorem }

\vspace{.5cm}

{\bf Enrique \'Alvarez $\dagger$ and C\'esar G\'omez}
\footnote{E-mail: {\tt enrique.alvarez, cesar.comez
@uam.es}}

\vspace{.3cm}

\vskip 0.4cm

{\it $\dagger$ Theory Division, CERN,1211 Geneva 23, Switzerland,\\
 Instituto de F\'{\i}sica Te\'orica, C-XVI,
\footnote{Unidad de Investigaci\'on Asociada
  al Centro de F\'{\i}sica Miguel Catal\'an (C.S.I.C.)}
and  Departamento de F\'{\i}sica Te\'orica, C-XI,\\
  Universidad Aut\'onoma de Madrid 
  E-28049-Madrid, Spain }

\vskip 0.2cm

\vskip 1cm


{\bf Abstract}

\end{center}

\begin{quote}
The equivalence between the holographic renormalization 
group and the soft dilaton theorem  
is shown for a class of warpped metrics solutions of the string
beta function equations for the bosonic string.  
  
\end{quote}


\newpage

\setcounter{page}{1}
\setcounter{footnote}{1}
\newpage


A renormalization group interpretation of the holographic map
(\cite{maldacena}\cite{gubser}\cite{witten}) between
quantum field theories (QFT) in four dimensions and classical gravity
in five-dimensional space-time 
has
been the subject of a bunch of recent papers
(\cite{akhmedov}\cite{alvarezgomez1}\cite{alvarezgomez2}\cite{alvarezgomez3}
\cite{alvarezgomez4}\cite{balasubramanian}\cite{freedman}\cite{girardello}
\cite{porrati}\cite{boer}\cite{verlindes}). 
The main idea underlying the {\em
  renormalization group approach} (RGA) to holography stems from the
interpretation of the running of the coupling constants of the four
dimensional QFT as Einstein's equations of five-dimensional gravity
plus matter. This approach rests on the holographic map relating QFT
couplings with boundary values of five-dimensional background fields.
\par
Let $S(g_{\m\n},\Phi_i)$ stand for the classical Einstein action evaluated on 
shell, at the points $g_{\m\n}(\bar{x},\rho),\Phi_i(\bar{x},\rho)$
solutions of the classical equations of motion with fixed boundary conditions,
$g_{\m\n}(\bar{x})\equiv g_{\m\n}(\bar{x},\rho=\infty);\Phi_i(\bar{x})
\equiv\Phi_i(\bar{x},
\rho=\infty)$ at $\rho=\infty$. The space-time background metric preserving
four dimensional Poincare invariance would be given by
\be
ds^2 = g_{\m\n}(\bar{x},\rho)dx^{\m}dx^{\n}+d\rho^2
\ee
The holographic map (\cite{maldacena}\cite{gubser}\cite{witten})
dictates that QFT
correlators are given in the large $N$ limit 
in terms of the classical action $S(g_{\m\n},\Phi_i)$ through:
\be
<\Theta_{i_1}(\bar{x}_1),\ldots \Theta_{i_n}(\bar{x}_n)>=
\frac{\delta^n S(g_{\m\n},\Phi_i)}{\delta\Phi_{i_1}(x_{i_1})\ldots
 \delta\Phi_{i_n}(x_{i_n})}|_{\Phi_i=\Phi_i^{(0)}}
\ee
  
where the $\Phi_i^{(0)}$ parametrize the particular QFT under consideration,
and where the correspondence
\be
\Phi_i\rightarrow\Theta_i
\ee
between five dimensional fields and four dimensional local observables
has been employed.
\par
The five dimensional bulk action $S(g_{\m\n},\Phi_i)$ is infrared (IR)
divergent (due to the contributions of the boundary $\rho=\infty$).
This corresponds to QFT ultraviolet (UV) infinities, and is the
essential ingredient of the IR/UV connection (\cite{susskindwitten}).A
regulator can be easily introduced, for example, by cutting down the
region $\rho >\Lambda $ (cf.\cite{henningson}\cite{boer}\cite{verlindes}), and
be used to define , in Wilsonian sense, correlators at the scale $\Lambda$.
Demanding that physical observables should be independent of the
particular value chosen for $\Lambda$ then implements the
Callan-Symanzik (CS) equations of the corresponding QFT. From the
five-dimensional bulk point of view, this is equivalent to demanding
invariance under translations in the holographic direction, 
$\rho\rightarrow\rho+a$. In (\cite{boer}) the CS equation was thus recovered
from the Hamilton constraint $H=0$ corresponding to the {\em holographic 
time} $\rho$ \footnote{  A Hamilton-Jacobi approach to the renormalization 
group for non critical strings on cosmological backgrounds was also
pointed out by Polyakov in reference (\cite{polyakov0})}
\par
Given a generic metric where four dimensions are fibered in a 
nontrivial way over the holographic coordinate, such as:
\be
ds^2 = a^2 (\rho) d\vec{x}_4^2 + d\rho^2,
\ee
where $d\vec{x}_4^2$ stands for the flat minkowskian metric,
the holographic version of Callan Symanzik equation for dimensionless 
physical quantities $A$, is ( \cite{boer}):
\be\label{vv}
(a\frac{\pd}{\pd a}- \b_i\frac{\pd}{\pd \Phi_i}) A =0
\ee
with 
\be\label{bi}
\b_i\equiv a\frac{\pd}{\pd a}\Phi_i.
\ee
 It is then plain that
the {\em warping} factor $a$ plays  precisely the r\^ole of the
renormalization group scale $\m$.\footnote{For an early version of this
identification, based on Polyakov's ideas for the confining 
string description of pure Yang-Mills (\cite{polyakov1}\cite{polyakov2}
\cite{polyakov3}), see \cite{alvarezgomez2}.}
\par
The purpose of this letter is to relate equation (\ref{vv}) with the string
soft dilaton theorem \cite{ademollo}.  
\par
Let us consider the simplest
case, namely the open bosonic string.  Divergences of pure gluonic
one-loop amplitudes come from the boundary of the moduli space of the
annulus. It is a well known fact since the early days of string theory
that those divergences could be absorbed into a renormalization of both the
string coupling constant, $g$ and the string tension, $\a'$.
To be specific, the open string one loop ultraviolet divergences can be
represented as insertions of {\em closed} string dilaton vertex operators
evaluated at at zero momentum . This is the reason why
the infinities can be absorbed into changes of $\a'$, because those
correspond to dilaton insertions.
The full soft dilaton theorem can be summarized in the following equation
(\cite{ademollo})
\be\label{sdt}
(l_s\frac{\pd}{\pd l_s} - \frac{(d-2)g}{4}\frac{\pd}{\pd g})
A(\kappa_1\ldots\kappa_n)=lim_ {p\rightarrow 0}A(p;\kappa_1\ldots\kappa_n)
J\Delta(p).
\ee
where $\a'\equiv l_s^2$, $d$ is the space-time dimension in which the gluonic
momenta, $\kappa_1\ldots\kappa_n$ can lie, and in the right 
hand side $J$ is
zero momentum dilaton tadpole and $\Delta(p)$ the dilaton propagator. The
coupling $g$ is the open string coupling constant.
\par
It is perhaps worth stressing that the validity of equation 
(\ref{sdt}) \cite{polyakov1}\cite{polyakov2}does not depend 
on working in the critical dimension.
\par
When the string background is a consistent solution of the sigma model
conformal invariance conditions (\cite{callan},\cite{curci}) dilaton tadpoles
should vanish, and the equation becomes:
\be
(l_s\frac{\pd}{\pd l_s} - \frac{(d-2)g}{4}\frac{\pd}{\pd g})
A(\kappa_1\ldots\kappa_n)= 0,
\ee
which when $d=4$ will be taken as our starting point. Let us evaluate it for
a conformally invariant string background, i.e., to lowest order in $\a'$, 
a solution of the equations:
\bea\label{background}
R_{\m\n}-\nabla_{\m}\nabla_{\n}\Phi = 0\nonumber\\
\nabla^2\Phi +(\nabla\Phi)^2 = 0
\eea
Based on the soft dilaton theorem, the following solution was derived in
 (\cite{alvarezgomez3})
\bea\label{dilaton}
ds^2=\frac{\rho}{l_c} d\vec{x}_4^2 + d\rho^2\nonumber\\
\Phi(\rho)=- log \rho
\eea
In \cite{alvarezgomez3}\cite{alvarezgomez4} has been argued that this metric
is in some sense a {\em universal} string background canonically associated
to  gauge quantum field theories. Using now the specific dilaton dependence
as determined in (\ref{dilaton}) 
in equation (\ref{bi}) yields:
\be
\b_{\Phi} = a\frac{\pd \Phi}{\pd a} = - 2
\ee
which turns the  holographic renormalization group equation (\ref{vv}) into:
\be\label{bvv}
(a\frac{\pd}{\pd a} + 2 \frac{\pd}{\pd\Phi}) A =0.
\ee
Using now the well-known relationship between the string coupling
and the dilaton background namely,
\be
g\equiv e^{-\Phi/2},
\ee
and interpreting $\rho$ in the metric (\ref{dilaton}) as a four-dimensional
running string tension (\cite{polyakov3}) easily yields from (\ref{bvv})
\be\label{eq}
(l_s\frac{\pd}{\pd l_s} - \frac{(d-2)g}{4}\frac{\pd}{\pd g})A = 0
\ee
which is {\em exactly} the four dimensional soft dilaton theorem.
\par
Please note that from the present point of view the dynamics underlying the 
Callan-Symanzik equation is just the conditions for the conformal invariance
of the two-dimensional string sigma model {\em and} the fact that open 
string one loop divergences just happen to be represented by insertions of 
closed string
vertex operators.
\par
As a general comment, it is plain that the open/closed interelationship
(as encoded in the soft dilaton theorem and otherwise) seems to underlie
the whole holographic interpretation of 
gauge theories.
\par
Four dimensional conformally invariant field theories , defined by $\b_i =0$,
do not renormalize the string coupling constant. Both this fact and 
the numerical value $(0)$ of the second member of the background equation
\footnote{Which in turn uniquely selects $d=4$.}
(\ref{background}) are related to poorly understood (closed) tachyon condensates and the Liouville nature of the holographic coordinate.
Finally we address the attention of the reader to the peculiar 
(from the renormalization group  viewpoint) 
{\em minus sign} both
in the holographic renormalization group (equation \ref{vv}) and in 
the soft dilaton theorem equation (\ref{sdt}). From the Hamilton-Jacobi point
of view \cite{boer} this sign could be related with the wrong sign
of the conformal factor in Einstein action while in the context
of the soft dilaton theorem the sign is related to invariance of
the gravitational constant $\kappa$ with respect to shifts of  the
dilaton field. The sign in (\ref{eq}) depends on the particular identification
between $\rho$ and $l_s$; we have used the dimension 
of $\rho$ as defined by the metric
(\ref{dilaton}).

\section*{Acknowledgments}
This work has been supported by the
European Union TMR programs FMRX-CT96-0012 {\sl Integrability,
  Non-perturbative Effects, and Symmetry in Quantum Field Theory}
and  ERBFMRX-CT96-0090 {\sl 
Beyond the Standard model} as well as
by the Spanish grants AEN96-1655 and AEN96-1664.


\appendix


               \
\end{document}